\documentclass[aps,prb,preprint,superscriptaddress]{revtex4-2}
\usepackage{csquotes}
\usepackage{amsmath,bm}
\usepackage{graphicx}
\begin{document}
\title{Observing the influence of second harmonic tangential surface source in gold plasmonic nanostructures}
\author{Sandy Mathew}
\affiliation{Université Grenoble Alpes, Grenoble-INP, Institut Néel, CNRS UPR-2940, 38000 Grenoble, France} 
\author{ Maëliss Ethis de Corny}
\affiliation{Université Grenoble Alpes, Grenoble-INP, Institut Néel, CNRS UPR-2940, 38000 Grenoble, France} 
\author{Nicolas Chauvet} 
\affiliation{Université Grenoble Alpes, Grenoble-INP, Institut Néel, CNRS UPR-2940, 38000 Grenoble, France} 
\author{Laureen Moreaud}
\affiliation{CEMES, CNRS UPR-8011, 31055 Toulouse, France}
\author{Erik Dujardin}
\affiliation{CEMES, CNRS UPR-8011, 31055 Toulouse, France}
\affiliation{ ICB, Université de Bourgogne, CNRS UMR-6303, 21078 Dijon, France} 
\author{Gilles Nogues}
\affiliation{Université Grenoble Alpes, Grenoble-INP, Institut Néel, CNRS UPR-2940, 38000 Grenoble, France} 
\author{Guillaume Bachelier}
\email{guillaume.bachelier@neel.cnrs.fr}
\affiliation{Université Grenoble Alpes, Grenoble-INP, Institut Néel, CNRS UPR-2940, 38000 Grenoble, France}
\begin{abstract}
The origin of the second harmonic generation (SHG) from plasmonic structures remains a subject of debate. Here, we investigate SHG from gold plasmonic nanostructures by scanning nanostructures at the single particle level to acquire 2D SHG maps which are qualitatively and quantitatively compared with numerically simulated SHG maps. This approach allows us to discriminate the role played by the surface and bulk contributions to SHG. Our findings reveal that the tangential surface source contribution, often neglected in the literature, plays a dominant role and is affected by surface roughness. This is evident when comparing the SHG maps obtained from identically sized gold nanoparticles that are fabricated using physical and chemical techniques. Additionally, the normal surface source is not found to significantly impact SHG in these configurations.
\end{abstract}
\maketitle
\clearpage
\section{Introduction}\label{Introduction}
Second harmonic generation (SHG) in metals has been a long-standing subject of fundamental as well as practical interest  spanning decades \cite{Jha,Agarwal1982, Kauranen}. This interest was fuelled by the presence of important features related to plasmonic effect that allows to enhance second-harmonic intensity and surface sensitivity \cite{Chen81, Hubert2007}. Indeed, under the electric dipole approximation, SHG is forbidden in centrosymmetric plasmonic metals \cite{Butet}. However, higher order effects arising from electric quadrupole and magnetic dipole lead to a second harmonic response from metals that have been described by several theoretical approaches \cite{Butet}. These effects have been naturally included in the free-electron models to describe SHG from the bulk of metals \cite{Bloembergenshg,Jha3,Chang}. On the other hand, description of SHG at the surface requires taking into account centrosymmetry breaking and non-local effects due to surface current density variation that extends to only few Fermi wavelengths \cite{Rudnick,Sipe}. Hydrodynamic model has therefore found some success due its simplicity whereby these effects are taken into account through convective acceleration, Lorentz force and quantum pressure contributions \cite{Sipe,Corvi,Krasavin}. Works involving numerical simulations using finite-difference time-domain techniques have also been applied to the hydrodynamic model \cite{Zeng2009, Liu2010}. In an effort to further accurately describe the electron density profile at the surface, density functional theory \cite{Weber,Liebsch} and quantum hydrodynamic model \cite{Schaich,Cirpendry2013,Khalid2020} have also been proposed.   \\ 
The macroscopic description of the second harmonic response considers surface and bulk contributions \cite{Wang, Krause} while more microscopic approaches describe it in terms of conduction and bound electron contributions \cite{Jha2,Scalora}. As the scientific interest shifted to nanoscale light-matter interaction, numerous works have been dedicated to understanding SHG from noble-metal nanoparticles and nanostructures \cite{Cong2024, Sugita2023}. Within this context, it has been shown that bulk source dominates SHG response in the case of gold split-ring resonators and T-shaped apertures \cite{Feth,Rahimi}. In a similar vein, theoretical study on 3D gold nanoantennas and experimental results obtained from spherical gold nanoparticles show that both bulk and surface sources of SHG can prevail depending on the experimental configuration \cite{Benedetti, Bachelier}. Whereas in other works, the normal component of the surface source
has been considered as the primary contribution in plasmonic nanoparticles \cite{Wunderlich2013, Timbrell, Cong2024}. A recent review \cite{Wang2023} states that \enquote{the surface SHG becomes more and
more predominant as the surface-to-volume ratio
increases. With respect to NPs, the bulk SHG can
be neglected due to the high surface-to-volume ratio}. Although one may intuitively expect to observe surface contributions dominating due to high surface to volume ratio, it has been shown that in the case of spherical nanoparticles size dependence cannot be used to distinguish the dominant nature of the nonlinear source contributions \cite{Bachelier}. Similarly, second harmonic response from asymmetric gold nanoantennas was simulated in \cite{Celebrano} by neglecting both the bulk source and the tangential component of surface source. \\ In this regard, the origin of SHG response in plasmonic nanostructures still appears to be a matter of debate. Here, we demonstrate that the bulk and the tangential surface sources plays a non-trivial role in the second harmonic response from gold nanostructures by comparing experimental and simulated far-field SHG intensity distribution. In particular, the interferences between different plasmonic modes determines the spatial distribution of SHG intensity recorded from individual nanostructures which is compared to numerical simulations. In contrast to the literature, we do not observe the dominant nature of the normal surface source, but report the first experimental evidence of the major role played by the second harmonic tangential surface source.
\section{Results and Discussion}
\subsection{Bulk and surface contributions}
\begin{figure}[h!]
\includegraphics[scale=0.7]{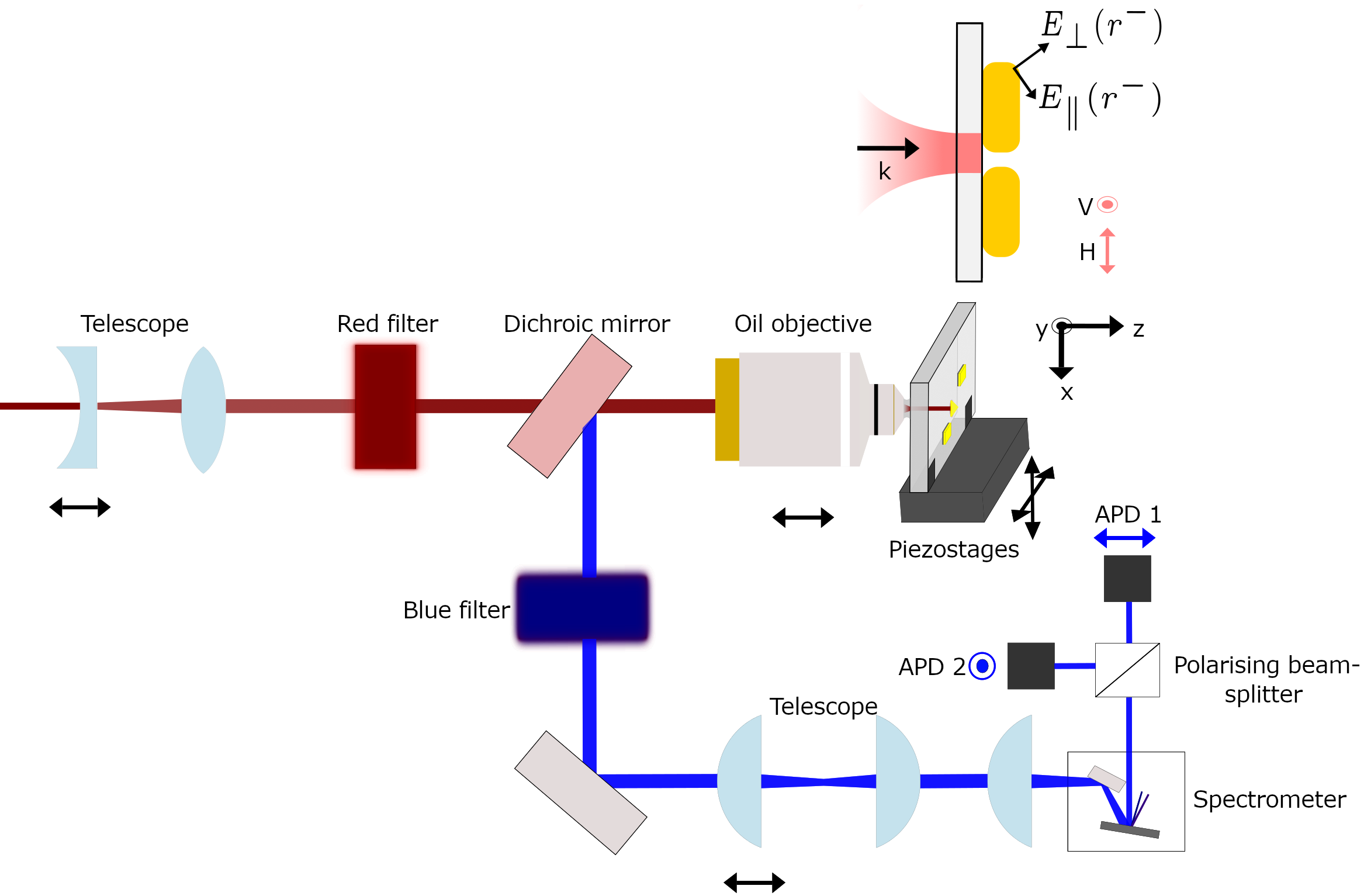}
\caption{The schematic of the optical setup used for acquiring the experimental SHG maps is shown in the figure. It consists of a femtosecond laser source which is focused first through a telescopic pair of lenses that allows to correct for chromatic aberrations of the oil objective lens. A similar optical system is used in the collection path for the same purpose. The red and blue filters represent band pass filters. The piezostages allow for nano-positioning. The polarising beam splitter after the spectrometer enables the collection of polarisation resolved SHG signal onto two avalanche photodiodes (shown as APD 1 and 2). The vectors (x,y,z) represent coordinate system with respect to the sample shown in the inset. $E_\parallel(r^-)$ and $E_\perp(r^-)$ represent parallel and perpendicular electric field components at the fundamental frequency just below the surface. V and H correspond to perpendicular and horizontal polarisation for the SHG signal. \label{setup}}
\end{figure}

The reduced form for the nonlinear second harmonic polarisation involving the non-local bulk contribution in the case of isotropic and centrosymmetric metal is given as \cite{Heinz1991, Bachelier}:
\begin{equation}
\begin{split}
\mathbf{P}_{bulk}(\mathbf{r},2\omega)&= \gamma_{bulk}\bm{\nabla}[\mathbf{E}(\mathbf{r},\omega)\cdot \mathbf{E}(\mathbf{r},\omega)],
\end{split}\label{Bulkpolarmain}
\end{equation}
where $\gamma_{bulk}$ is the bulk susceptibility and $\mathbf{E}$ is the electric field at the fundamental frequency. The non-local nature comes from the gradient, in other words, the induced polarisation is not related to the local value of the electric field but to gradient of its squared value. \\ The breaking of the centrosymmetry at a metal surface leads to the nonlinear second harmonic surface polarisation given as \cite{Heinz1991, Bachelier}:
\begin{figure}[h]
\begin{equation}
\begin{split}
\mathbf{P}_{surface}(\mathbf{r}^+,2\omega)&=\mathbf{P}_{\perp}(\mathbf{r}^+,2\omega)+\mathbf{P}_{\parallel}(\mathbf{r}^+,2\omega)\\&=\chi_{\perp\perp\perp}E_{\perp}(\mathbf{r}^-,\omega)E_{\perp}(\mathbf{r}^-,\omega)\textbf{e}_{\perp}\\&+\chi_{\parallel\parallel\perp}E_{\parallel}(\mathbf{r}^-,\omega)E_{\perp}(\mathbf{r}^-,\omega)\textbf{e}_{\parallel}\label{surfacepolarmain}
\end{split}
\end{equation}
\end{figure}
\\where $\perp$ and $\parallel$ indicate components perpendicular and tangential to the surface as shown in the schematic (Fig.~\ref{setup}) and $\textbf{e}_{\perp} $, $\textbf{e}_{\parallel}$ represents corresponding unit vectors.  The plus and minus signs represents the dielectric side and the metal side respectively. For isotropic and centrosymmetric materials the perpendicular and parallel components are mapped to the susceptibility tensor components as $ {\chi_{xxz}}^s={\chi_{xzx}}^s={\chi_{yyz}}^s={\chi_{yzy} }^s= \chi_{\parallel\parallel\perp}, {\chi_{zzz}}^s= \chi_{\perp\perp\perp}, {\chi_{zxx}}^s={\chi_{zyy}}^s=  \chi_{\perp\parallel\parallel}$ whereby \textit{z} is normal to the surface \cite{Wang}. ~In addition to the above mentioned second harmonic sources, there exists other components from the bulk \cite{Wang} written as $\beta\mathbf{E}(\mathbf{r},\omega)[\bm{\nabla}\cdot~\mathbf{E}(\mathbf{r},\omega)], \delta[\mathbf{E}(\mathbf{r},\omega)\cdot~\bm{\nabla}]\mathbf{E}(\mathbf{r},\omega)$ as well as from the surface expressed as $\chi_{\perp\parallel\parallel}E_{\parallel}(\mathbf{r}^-,\omega)E_{\parallel}(\mathbf{r}^-,\omega)$. While for homogeneous and isotropic materials the divergence of the field vanishes, others have been shown to be over one to two orders of magnitude smaller than the above sources which is too weak to be measurable in our case \cite{Wang}.
The bulk and surface contributions are given different weights through non-dimensional parameters namely \textit{a}, \textit{b} and \textit{d}, introduced by Rudnick and Stern into their associated susceptibility components \cite{Rudnick,Sipe}:
\begin{align}
\chi_{\perp\perp\perp}&=-\dfrac{a}{4}[\epsilon_r(\omega)-1]\dfrac{e\epsilon_0}{m\omega^2}\\
\chi_{\parallel\parallel\perp}&=-\dfrac{b}{2}[\epsilon_r(\omega)-1]\dfrac{e\epsilon_0}{m\omega^2}\\
\gamma_{bulk}&=-\dfrac{d}{8}[\epsilon_r(\omega)-1]\dfrac{e\epsilon_0}{m\omega^2}
\end{align}
In the above expressions, $\epsilon_r(\omega)$ is the dielectric function of the metal at frequency $\omega$ while \textit{e} and \textit{m} correspond to electron charge and mass. In the hydrodynamic model, their values are shown to be \textit{a}~=~\textit{d}~=~1 and \textit{b}~=~-1 \cite{Sipe}. The parameter \textit{a} has been shown to be frequency-dependent while \textit{b} is expected to be dependent on surface details due to the tangential component of the surface current and \textit{d} to be largely independent of either effects \cite{Sipe,Corvi,Weber}. The values set by the hydrodynamic model were later revised to include contribution from interband transitions in the case of gold spherical nanoparticles  by fitting experimental polarimetric data with computed electric fields so that \textit{a}~=~0.56-i0.25, \textit{b}~=~0.1 and \textit{d}~=~1 \cite{Bachelier}.\\
Finally it has been demonstrated that the non-local bulk contribution in Eq. \ref{Bulkpolarmain} cannot be separately measured as it appears in a linear combination with the normal component of the surface polarisation in Eq. \ref{surfacepolarmain} \cite{Sipe2}. More precisely, the non-local bulk contribution can be reformulated into an effective surface contribution as \cite{Maeliss, Sipe2}:
\begin{align}
\begin{split} \label{separablesources}
{P}_{\perp}(\mathbf{r}^+,2\omega)&=\big(\dfrac{\gamma_{bulk}}{\epsilon(2\omega)}+\chi_{\perp\parallel\parallel}\big)\mathbf{E}(\mathbf{r}^-,\omega)\cdot \mathbf{E}(\mathbf{r}^-,\omega) \\
&+\big(\chi_{\perp\perp\perp}-\chi_{\perp\parallel\parallel}\big){E}_{\perp}(\mathbf{r}^-,\omega){E}_{\perp}(\mathbf{r}^-,\omega)
\end{split}
\end{align}
However, as $\chi_{\perp\parallel\parallel}$ is expected to be vanishing \cite{Teplin2}, the effective polarisation can then be written as: 
\begin{align}
\begin{split} \label{separablesources2}
{P}_{\perp}(\mathbf{r}^+,2\omega)&=\dfrac{\gamma_{bulk}}{\epsilon(2\omega)}\mathbf{E}(\mathbf{r}^-,\omega)\cdot \mathbf{E}(\mathbf{r}^-,\omega) \\
&+\chi_{\perp\perp\perp}{E}_{\perp}(\mathbf{r}^-,\omega){E}_{\perp}(\mathbf{r}^-,\omega)
\end{split}
\end{align}
From the above expression one can see that the electric field dependence is not the same between surface and bulk contributions. This field dependence of the two sources can be used as a signature to discriminate between their respective contributions. This has already been demonstrated in the case of resonant single aluminium nanoantennas \cite{Maeliss}. 
\\In order to work at the single nanoparticle level, our experimental setup was designed to allow for diffraction-limited spatial resolution, nano-positioning of sample, low optical noise etc (Fig.~\ref{setup}). Excitation was performed with a Ti:Sapphire femtosecond laser of 100~fs pulse width with a repetition rate of 80~MHz was used. The average power used was $200~\mu W$.  The beam was directed to a band pass filter first (shown as red filter in Fig.~\ref{setup}) to remove any stray SHG emission in the path caused by the excitation beam. In order to achieve diffraction-limited resolution, the laser was focused through a high numerical aperture (x100, N.A = 1.3) oil objective and the optical signal was collected in reflection mode. Consequently, for an excitation wavelength of $~1000$~nm the spot size at excitation is $~385$~nm and $~192$~nm at the harmonic wavelength $(\lambda/(2\times N.A)$).  The sample substrate used was a Zeiss glass coverslip of thickness $170~\mu m$ hosting the nanostructures. The chromatic aberrations from the objective lens at the operating wavelengths were compensated by using telescopic lenses such that the focal planes overlap the sample surface. To scan the nanoparticle using the focused beam, the sample was placed on closed-loop piezoelectric stages of nanometric precision. The xy scan step size was 50nm. The SHG signal was separated from the input beam path by using a dichroic beam splitter and filtered using a band pass filter (350~nm – 520~nm shown as blue filter in Fig.~\ref{setup}) to transmit only the SHG wavelength corresponding to the excitation. The SHG signal was sent to a diffraction grating with 150 grooves/mm followed by a polarising beam splitter such that the signal is decomposed into two orthogonal polarization components, each detected by a dedicated detector. The detectors were low noise avalanche photodiodes having roughly 10 dark counts per second, shown as APD1 and APD2 in Fig.~\ref{setup}. To reduce the noise further an ultrafast acquisition card was synchronised with the laser pulses so that it was possible to discriminate between SHG signals and dark counts for each detector \cite{Maeliss}. 
\subsection{Finite element simulations}
Our approach to discriminate the sources is to simulate their respective contributions in a quantitative manner and to compare it with experimentally obtained response. The simulations help finding configurations where spatial decorrelation of the second harmonic source response becomes evident. \\ In order to achieve a quantitative comparison between experiment and simulation, the simulations were performed by combining analytical methods (Matlab) and finite element method (COMSOL) to take into account the experimental setup described previously \cite{Chauvet}. In the analytical step, propagation of the excitation laser field through an oil immersion microscope objective of N.A~=~1.3 is evaluated using a wavelet approach \cite{Novotny}. In a similar manner, the transmission of the field through oil and glass substrate hosting the nanoantennas is evaluated using the formalism described in \cite{Novotny}. The material properties of gold was taken from the literature \cite{Johnson1972}. The dielectric function in our simulations do not account for nonlocal effects but only includes linear optical properties. The nonlinear source terms are therefore explicitly included. The refractive index used for the glass substrate is fixed to n =1.518. As the experimental setup is conceived to work in reflection, the second harmonic signals generated by the nanoantennas are back-propagated, collected and transmitted taking into account all the aforementioned elements and finally radiated onto a detector in the far-field. Here again, we use directly the Green's tensor framework of \cite{Novotny} (see \cite{Laurent23} for more details). The electromagnetic field interaction with the nanoantennas is evaluated by solving Maxwell's equations within scattered field formulation using finite element method. The total field obtained at the fundamental frequency ($\omega$) is used to compute nonlinear source currents using Eq.\ref{Bulkpolarmain} and Eq. \ref{surfacepolarmain}. With the nonlinear source currents, the Maxwell's equations are solved at the harmonic frequency ($2\omega$) using a weak formulation as shown below. It is used to compute the electric field at the harmonic frequency to be radiated to the detectors  \cite{Chauvet}. 
\begin{equation}
\begin{split}
&\int -j\mu_0 2\omega\mathbf{J}_{surface}(\mathbf{r}^+,2\omega)\cdot \text{test}(\mathbf{E}(\mathbf{r}^+,2\omega)) dS +\int -j\mu_0 2\omega\mathbf{J}_{bulk}(\mathbf{r},2\omega)\cdot \text{test}(\mathbf{E}(\mathbf{r},2\omega)) dV
\label{weakform}
\end{split}
\end{equation}
The test function in the weak formulation is a sampling function used to reduce the smoothness requirement of strong form equations. It is the standard way of implementing weak form in Comsol. This framework is adopted since it allows evaluating surface current components of any orientation as well as it is possible to place polarisation sheet just outside the metal surface as required by the description provided earlier \cite{Sipe,NireekshanReddy17}.
Hence, this approach allows locating the surface polarisation sources as stated in Eq. \ref{surfacepolarmain}, i.e. just above the metal while the electric fields generating them are evaluated just inside.
\subsection{Interplay between bulk and tangential surface source in gold gap nanoantennas}
\begin{figure*}[h!]
\centering
\includegraphics[scale=1.9]{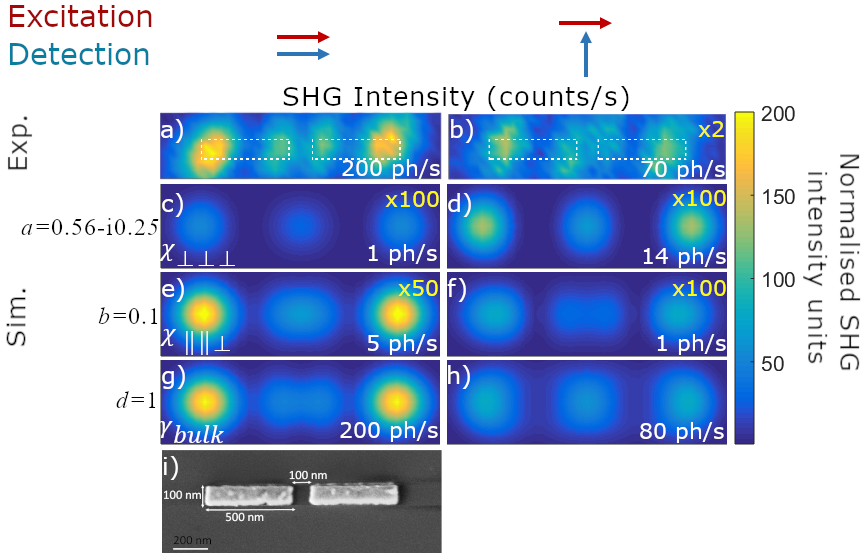}
\caption{A scanning electron microscope (SEM) image of a gap antenna of gold of size 500~nm separated by a gap of 100~nm is shown along with the simulated and experimental SHG maps obtained from the antenna structure when excited at 940~nm with the field polarised along the nanoantenna axis (red arrows). The collected SHG signal is polarisation resolved shown by the blue arrows. The simulated SHG source responses are indicated as $\chi_{\perp\perp\perp}$, $\chi_{\parallel\parallel\perp}$ and $\gamma_{bulk}$ representing normal surface source, tangential surface source and non-local bulk source along with their respective Rudnick and Stern parameters. The numbers in yellow in the inset of the figures represent multiplicative factors for better visibility of the SHG intensity distribution while those in white are the peak intensity values. \label{gapant}}
\end{figure*}
\begin{figure*}[h!]
\centering
\includegraphics[scale=1.9]{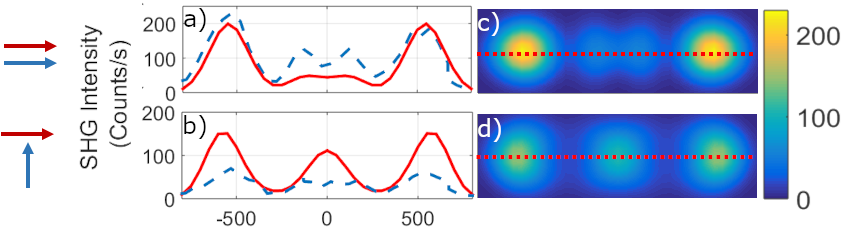}
\caption{ Intensity profiles of the coherent sum of sources and experimental profiles are shown in (a-b) corresponding to the two detection polarisation configurations. The colour in red represents simulated intensity profiles traced along the dotted line in red shown on the simulated SHG maps (c-d). The dotted profiles in colour blue represents experimental intensity profiles. The weights given by the Rudnick and Stern parameters are a=0.56-0.25i, b=0.1, d=1.\label{gapantcoh}}
\end{figure*}

 
As mentioned before, due to different dependencies of the nonlinear sources on the fundamental field components, the second harmonic responses from the source currents can be identified in specific configurations: The nanostructures are scanned in the focal plane of the microscope objective so that the varying weight of excited plasmonic modes leads to distinct interference patterns in SHG maps. \\Here, we present first a configuration consisting of gold gap nanoantennas. Two rectangular nanoantennas of 500~nm in length, 100~nm in width and 35~nm in thickness spaced by a 100~nm gap are investigated. They were fabricated using electron beam lithography followed by electron gun evaporation. In Fig.~\ref{gapant}, experimental SHG maps and simulated SHG maps corresponding to the three nonlinear sources ($\chi_{\perp\perp\perp}$, $\chi_{\parallel\parallel\perp}$, $\gamma_{bulk}$) with Rudnick and Stern parameters \textit{a}~=~0.56-i0.25, \textit{b}~=~0.1 and \textit{d}~=~1 \cite{Bachelier} are presented. The excitation wavelength is 940~nm and the excitation polarisation (red arrows) is along the antenna long axis. We only show here the SHG maps for horizontal excitation polarisation since the corresponding orthogonal configuration leads to significantly lower signal to noise ratio. The polarisation resolved SHG shown by the blue arrows are measured simultaneously by two the two APD detectors as shown in the Fig.~\ref{setup} to ensure position correlation between the two maps. All the simulated maps were renormalized by a common factor. It was determined by the highest intensity obtained from the non-local bulk source, and further scaled by a common multiplicative factor corresponding to the highest experimentally measured counts per second. Care must be taken in the interpretation of the SHG intensity distribution presented in the simulated as well as experimental maps.  As mentioned earlier, these maps are obtained by collecting SHG intensity as a function of excitation focus position. Thus, these maps do not display a given plasmonic mode rather it represents the product of the efficiency of excitation of the plasmons and the efficiency of SHG collected from the excitation position, corresponding to a diffraction limited spot at the harmonic wavelength.\\ A few conclusions can be drawn by a simple and direct observation from these maps: (i) The bulk contribution is dominant based on the magnitude of the normalised intensity in the simulated maps \cite{Feth,Rahimi}, (ii) The SHG intensity in the experimental maps (Fig.~\ref{gapant}~(a-b)) is higher when the detection polarisation is aligned parallel to the excitation polarisation which is in contradiction to the behaviour of $\chi_{\perp\perp\perp}$ surface contribution (Fig.~\ref{gapant}~(c-d)). Thus, it can be ruled out as the dominant source, (iii) While the simulated maps suggests that the bulk contribution is dominant regardless of the detection polarisation. This can be noted clearly in Fig.~\ref{gapantcoh} which represents coherent sum of the three sources. However, the intensity distribution at the gap from $\chi_{\parallel\parallel\perp}$ source in the crossed detection polarisation (Fig.~\ref{gapant}~(f)) in fact matches better with the experimental map in Fig.~\ref{gapant}~(b). As the intensity obtained from the bulk source is much larger than the tangential surface source (due to the parameter $b~=~0.1$; see Fig.~\ref{gapant}~(e-f)), it suggests that the ratio of the associated weights of tangential surface source and non-local bulk source given by the Rudnick and Stern parameters (\textit{b}/\textit{d}) is largely underestimated. In other words, \textit{b} should be higher. This discrepancy in the relative weight of the bulk source contribution has already been pointed out \cite{Maeliss}. This simple analysis demonstrates that both non-local bulk source $\gamma_{bulk}$ and tangential surface source $\chi_{\parallel\parallel\perp}$ are equally important to the SHG response in contrast to the reported literature \cite{Timbrell, Wang2023}. The condition under which bulk contribution can become dominant was explored in \cite{Rahimi} such that maximising electric and magnetic components at the same spatial location can outweigh its contribution to surface sources. 
\\
 As the gap nanoantennas investigated were fabricated by electron beam lithography, they present a grainy surface which is not ideal to analyse the effect of tangential surface source beyond the discussion presented above. As the \textit{b} parameter corresponding to the $\chi_{\parallel\parallel\perp}$ source is expected to be sensitive to the roughness of surface \cite{Rudnick, Sipe}, varying the surface rugosity of the nanostructures by means of different fabrication methods should affect the SHG contribution from this source. 
\subsection{Dominant tangential surface source and non-local bulk source in smooth and rough nanoprisms}
\begin{figure*}[h]
\centering
\includegraphics[scale=0.6]{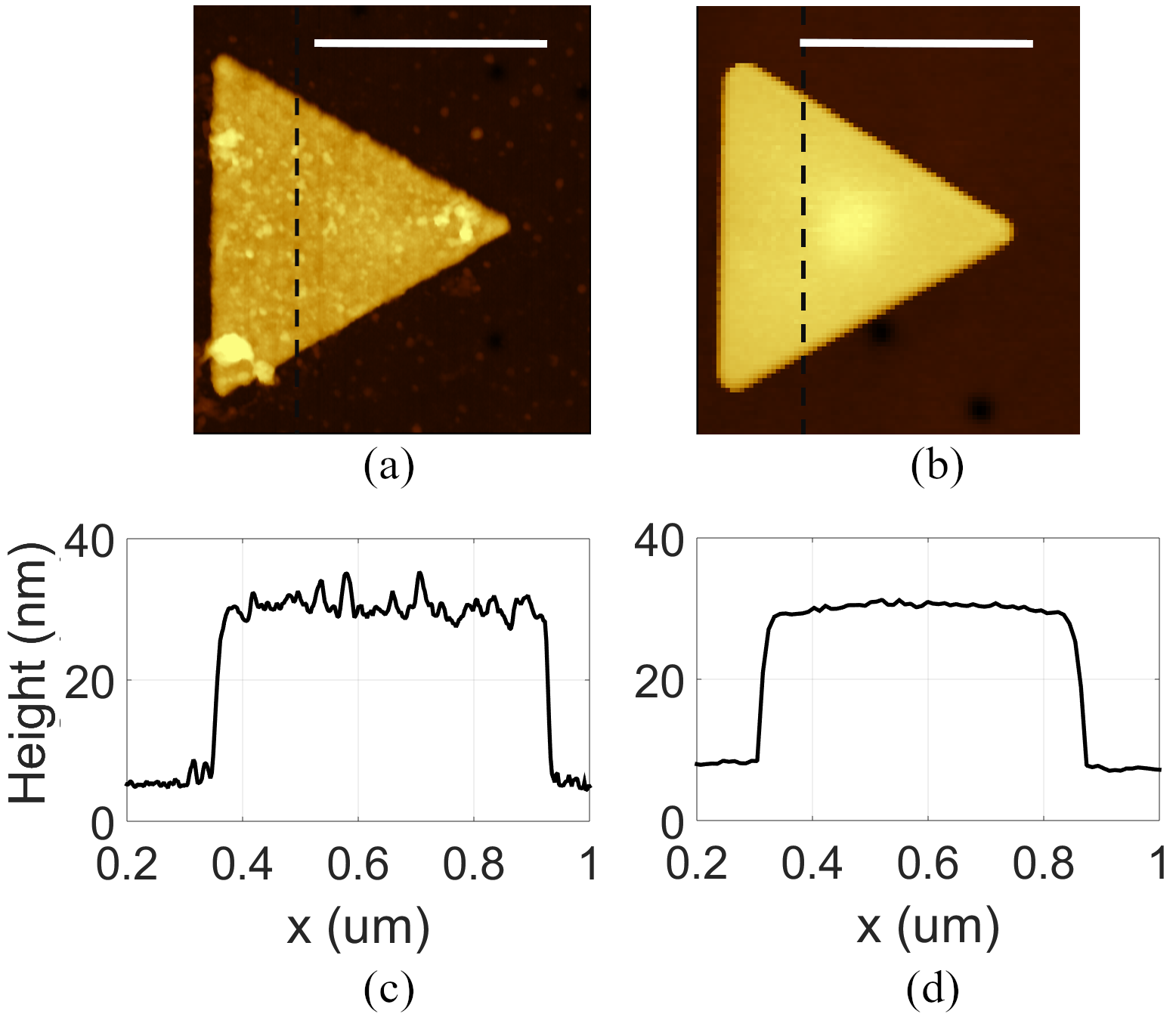}
\caption{AFM images of (a) a nanoprism fabricated via electron beam lithography along with its surface profile in (c) and (b) a crystalline prism with its surface profile in (d). Their average edge size is measured to be 757~nm and of thickness 22~nm. Scale bar 500~nm. \label{afmprism}}
\end{figure*}
\begin{figure*}
\centering
\includegraphics[scale=1.8]{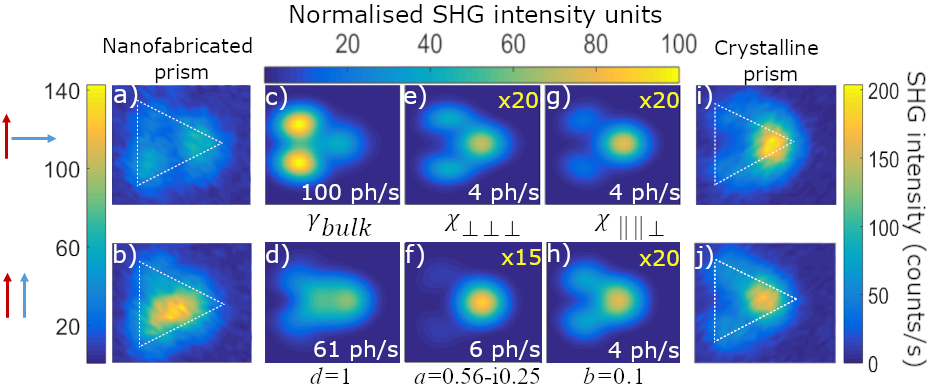}
\caption{Polarisation resolved experimental SHG maps of nanofabricated prism (a-b) and crystalline prism (i-j) corresponding to an excitation wavelength of 1000 nm polarised normal to the horizontal axis of the nanoprisms compared with simulated second harmonic source contributions (c-h) of the same configuration. The Rudnick and Stern parameters for the sources are presented at bottom of panels (d,f,h). The intensity of the maps corresponding to $\chi_{\perp\perp\perp}$ (e-f) and $\chi_{\parallel\parallel\perp}$ (g-h) are multiplied by a factor 20 for better visibility. The numbers in white in the inset are the peak intensity values. The dotted white lines on the experimental maps are guide to the eyes. \label{sources}}
\end{figure*}

We perform similar SHG analysis as above on gold nanoprisms of comparable geometry and dimensions either fabricated by electron beam lithography lift-off or by colloidal chemical synthesis.
The chemically synthesized nanoprisms
are crystalline and have an atomically smooth surface \cite{Viarbitskaya2013}. The difference between the two types of nanoprisms is clearly seen from the surface profiles shown in Fig.~\ref{afmprism}. Their dimensions are estimated to be 757~nm in edge length and 22~nm in thickness based on AFM and SEM images in both cases. 
\begin{figure*}[h]
\centering
\includegraphics[scale=1.7]{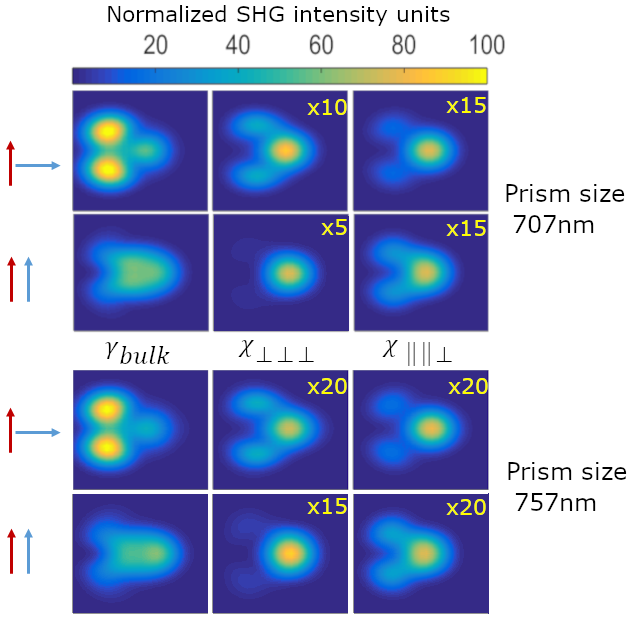}
\caption{ Comparison between simulated SHG source responses from nanoprisms of size 707~nm and 757~nm. \label{prismcomp_rev2}}
\end{figure*} 
In order to verify the effect of slight differences in the geometrical characteristics between the two nanoprisms, responses from nanoprisms of size 707~nm and 757~nm were simulated. Their corresponding maps are shown in Fig.~\ref{prismcomp_rev2}. In this case, we observe that there is only a little appreciable size-related effect. This is consistent with the dark field spectral analysis of nanoprisms of similar sizes showing that a size increase of at least 100~nm is necessary to accommodate a new set of plasmonic modes \cite{Erik2019}. Therefore, for our purpose, it is reasonable to consider that the crystalline and nanofabricated prisms (Fig.~\ref{afmprism}) are effectively identical. The SHG maps acquired on these nanoprisms are shown in Fig.~\ref{sources}~(a-b, i-j) along with the simulated source contributions (Fig.~\ref{sources}~(c-h)). The simulated SHG maps for the three sources are renormalised by a factor 100 based on the highest intensity counts obtained corresponding to the non-local bulk contribution (Fig.~\ref{sources}~(c)). 
\subsubsection{Nanofabricated nanoprism}
As it can be seen from the maps in Fig.~\ref{sources}, the excited plasmon modes in the nanoprisms lead to richer and more complex second harmonic responses than for the nano-gap antennas. Nevertheless, it is clear that SHG maps recorded for the two similarly sized gold nanoprisms under identical experimental conditions are drastically different suggesting that the underlying dominant source is different. 
\begin{figure*}[h!]
\centering
\includegraphics[scale=1.9]{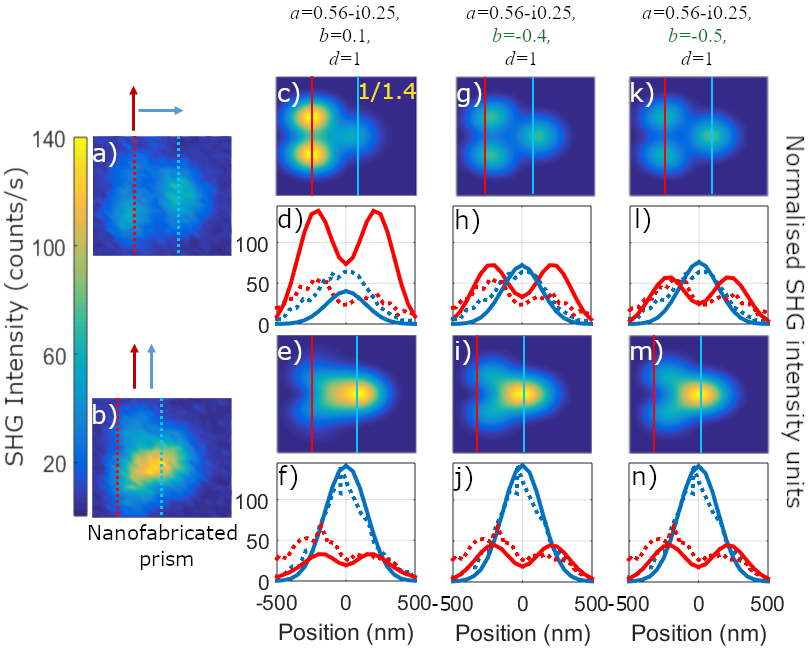}
\caption{SHG maps from lithograph prism (a-b) compared with the coherent sum of the three second harmonic sources  represented by their Rudnick and Stern parameter \textit{a,b,d} for parallel and crossed detection polarisations and their respective intensity profiles. The parameter \textit{b} is varied in (g-n) by the values shown in the header of each column. The dotted lines represents experimental profiles traced over the positions shown by the dotted lines drawn on the experimental SHG maps. The solid lines represents intensity profiles from the simulated SHG maps. The solid and dotted lines are colour coded based on the position over which intensity profile is traced. The intensity in (c) is reduced by a factor 1.4 for the purpose of normalisation.  \label{lithocoherent}}
\end{figure*}
\\ Three important observations can be drawn directly from Fig.~\ref{sources}: (i) As for the gap antennas, the intensity corresponding to the non-local bulk source $\gamma_{bulk}$ (Fig. \ref{sources} (c-d)) largely dominates the simulated SHG response compared to the surface sources, (ii)  The maps of the nanofabricated prism (Fig. \ref{sources} (a-b)) have qualitatively similar intensity distributions when compared with $\gamma_{bulk}$. For instance, the experimental map for the parallel detection polarisation with respect to excitation (Fig. \ref{sources} (b)) has a broad distribution centred on the nanoprism which is more comparable to $\gamma_{bulk}$ than for the two other sources. However, its magnitude of intensity contradicts the experimental map suggesting that $\gamma_{bulk}$ cannot be the only dominant source. (iii) In the case of parallel detection polarisation for $\chi_{\perp\perp\perp}$ (Fig.~\ref{sources}~(f)), the intensity distribution on the left side vanishes entirely compared to (Fig.~\ref{sources}~(e)) leading to a single prominent intensity lobe. On the contrary for $\chi_{\parallel\parallel\perp}$ source (Fig.~\ref{sources}~(g-h)), a spatially uniform increase in intensity is observed between the respective detection polarisations. Based on the variation in the magnitude of intensity of experimental maps (Fig.~\ref{sources}~(a-b)), this is qualitatively better matched by the response from $\chi_{\parallel\parallel\perp}$.
\\Hence consistent with the previous observation on gap nanoantennas, it
appears that the SHG response of the nanofabricated prism also arises from
a mixing between $\gamma_{bulk}$ and $\chi_{\parallel\parallel\perp}$ while $\chi_{\perp\perp\perp}$ can be ruled out. To quantitatively validate this premise, we show in Fig. \ref{lithocoherent} the intensity distribution computed from a coherent sum of the three sources when the contribution from $\chi_{\parallel\parallel\perp}$ is varied (b~=~0.1, -0.4, -0.5). The intensity profiles are normalised with respect to the maps obtained from the detection polarisation parallel to the excitation polarisation. In Fig.~\ref{lithocoherent}~(c-f), the values of \textit{a, b, d} parameters are taken from the literature \cite{Bachelier}. The profiles are taken from the intensity lobes at the left and right apex. Fig.~\ref{lithocoherent}~(d) clearly demonstrates that the ratio between intensity profiles from the simulated map considerably overshoots the experimental intensity profile. On the other hand, as the value of $|b|$ is increased  (Fig.~\ref{lithocoherent}~(g-h) and (k-l)), the SHG maps progressively coincide with  the experimental map. In particular, the intensity profiles quantitatively match when $|b|~=~0.5$ with a perfect agreement for the polarisation ratios (Fig.~\ref{lithocoherent}~(l and n)). This is a clear evidence of the non-trivial role played by the $\chi_{\parallel\parallel\perp}$ source. 
\subsubsection{Crystalline nanoprism}
\begin{figure*}[h!]
\centering

\includegraphics[scale=1.9]{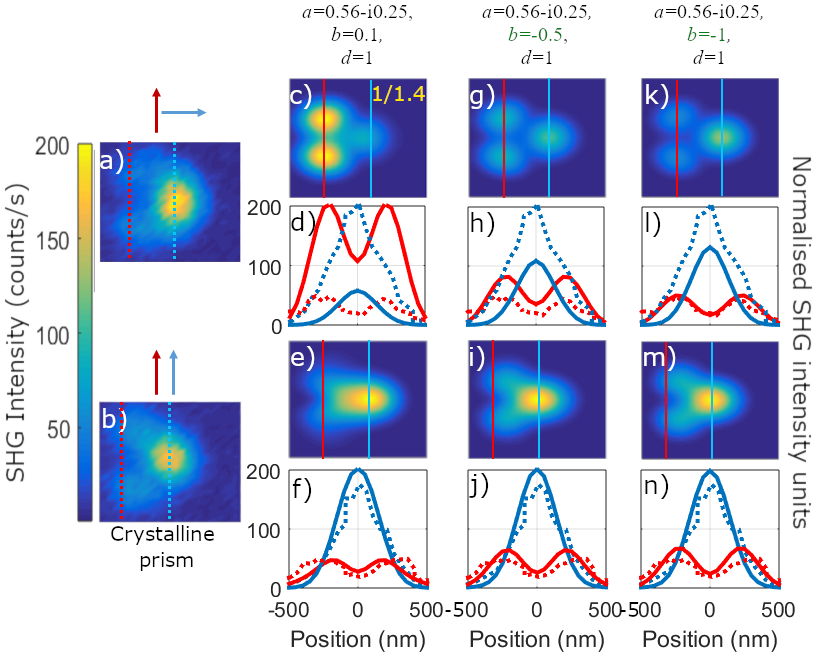}
\caption{SHG maps from monocrystalline prism (a-b) compared with the coherent sum of the three second harmonic sources  represented by their Rudnick and Stern parameter \textit{a,b,d} for parallel and crossed detection polarisations and their respective intensity profiles. The parameter \textit{b} is varied in (g-n) by the values shown in the header of each column. The dotted lines represents experimental profiles traced over the positions shown by the dotted lines drawn on the experimental SHG maps. The solid lines represents intensity profiles from the simulated SHG maps. The solid and dotted lines are colour coded based on the position over which intensity profile is traced. The intensity in (c) is reduced by a factor 1.4 for the purpose of normalisation.  \label{chemsourceprofile}}
\end{figure*}
\begin{figure*}[h]
\centering
\includegraphics[scale=1.9]{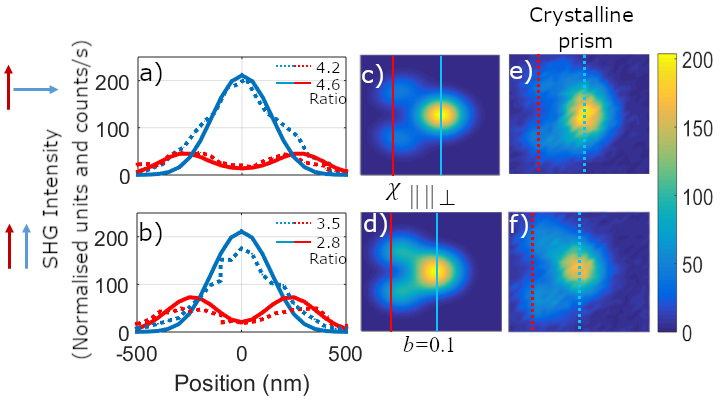}
\caption{The intensity profiles of the $\chi_{\parallel\parallel\perp}$ source are compared with the profiles obtained from the SHG maps of the crystalline prism. The solid and dotted lines correspond to simulated and experimental profiles respectively. The quantity in the inset of the plots represents the ratio between the highest intensities corresponding to the profiles traced on the experimental and simulated maps.  \label{chemsource}}
\end{figure*}

The SHG intensity distribution from the crystalline prism shown in Fig.~\ref{sources}~(i and j) is clearly inconsistent with the simulated $\gamma_{bulk}$ response for the same detection configurations (Fig.~\ref{sources}~(c and d)), thus ruling out $\gamma_{bulk}$ as the main contribution. For the $\chi_{\perp\perp\perp}$ contribution (Fig.~\ref{sources}~(e-f)), the intensity of the two lobes on the left side diminishes when the detection polarisation is parallel to the excitation while the intensity of the right apex increases. None of these two variations are observed in the experimental maps (Fig.~\ref{sources}~(i-j)) allowing here again to rule out $\chi_{\perp\perp\perp}$. On the contrary, for $\chi_{\parallel\parallel\perp}$ (Fig.~\ref{sources}~(g-h)), the intensity on the two left lobes increases while the intensity of the right apex remains unchanged. Qualitatively, the overall spatial distribution and intensity variation of  $\chi_{\parallel\parallel\perp}$ upon changing the detection polarisation matches with the maps obtained from the crystalline prism (Fig.~\ref{sources}~(i-j)). Figure~\ref{chemsourceprofile} shows the coherent sum of the SHG sources and their respective intensity profiles for $b~=~0.1, -0.5, -1$ compared with the experimental maps. Once again, the normalisation is made with respect to the maps from the detection polarisation parallel to the excitation (Fig.~\ref{chemsourceprofile}~(e, i, m)). In this case too, the maps and profiles in Fig.~\ref{chemsourceprofile}~(c-f), which are computed with values for the $a, b, d$ parameters taken from the literature \cite{Bachelier}, do not account well for the experimental data. As the value of \textit{b} is increased to $|b|~=~0.5$ (Fig.~\ref{chemsourceprofile}~(h)) the ratio between the two experimental profiles is drastically changed but remains different from the experimental data, in particular for the right apex in the orthogonal configuration.
When we consider the theoretical maximum for \textit{b}~($b~=-1$) from the hydrodynamic model in Fig.~\ref{chemsourceprofile}~(k-n) \cite{Liebsch89} the simulated intensity profiles are more consistent with experimental profiles. Note that this case corresponds to the theoretical value for a perfectly smooth and flat surface \cite{Sipe}. Yet, the ratio between the intensity profiles in Fig.~\ref{chemsourceprofile}~(l) does not fully agree and suggest that $|b|$ should be further increased. Hence, we consider in Fig.~\ref{chemsource} a pure $\chi_{\parallel\parallel\perp}$ contribution. Here, the maps are in better agreement indicated by the intensity profiles (Fig.~\ref{chemsource}~(a) and (b)). This is further confirmed by the intensity ratio calculated from the intense right apex and the dimmer left apexes in the experimental and simulated maps, as given in the insets of Fig.~\ref{chemsource}~(a) and (b), showing good agreement. This strongly suggest that $\chi_{\parallel\parallel\perp}$ is the dominant SHG source in crystalline prism. To the best of our knowledge, this is the first reported evidence of its dominant character. A more direct way to validate this parameter would be valuable, for instance, using an experimental approach as demonstrated in \cite{Wang}.    
\\To conclude, by comparing the SHG response obtained from single gold nanostructures, notably gap nanoantennas and nanoprisms, and numerical simulations we were able to demonstrate the nonlinear source that has a dominant character. In the case of nanoprisms, we provide the first reported evidence of the dominant nature of tangential surface source which is usually neglected in the literature. Furthermore, we also report cases where an interplay between the contributions from the non-local bulk source and tangential surface source can occur showing that they have a non-trivial contribution to SHG. The observation that the dominant nature of the tangential surface source  increases with respect to flatness or smoothness of a surface implies that it can be potentially used as a probe to assess surface quality. Finally, the contribution from $\chi_{\perp\perp\perp}$ source was found to disagree with experimental data irrespective of the antenna shape investigated here and surface roughness.

\begin{acknowledgments}
The authors acknowledge the financial support from the Agence Nationale de la Recherche, France, through the TWIN  project (Grant No. ANR-14-CE26-0001-01-TWIN), PlaCoRe and HybNap projects (Grants ANR-13-BS10-0007-PLACORE and ANR-16-CE09-0027-HybNaP) as well as the Université Grenoble Alpes, France, for the Chaire IUA award to G. Bachelier and Ph.D. grant to S. Mathew. 
\end{acknowledgments}

\bibliographystyle{apsrev4-2}

%

\end{document}